\documentclass[runningheads]{llncs}
\usepackage{graphicx}
\begin{document}
\title{How to Hijack Twitter:\\ Online Polarisation Strategies of Germany's political far-right}

\titlerunning{How to Hijack Twitter}
%
\author{Philipp Darius\inst{1} \and
Fabian Stephany\inst{2,3}\thanks{Corresponding author: \email{fabian.stephany@oii.ox.ac.uk}}
}
\authorrunning{Darius and Stephany}

%
\institute{Hertie School of Governance, Berlin, \and
Oxford Internet Institute, University of Oxford, \and
Humboldt Institute for Internet and Society, Berlin.
\\
}

\maketitle              
\begin{abstract}
With a network approach, we examine the case of the German far-right party \textit{Alternative f\"ur Deutschland} (AfD) and their potential use of a \textit{"hashjacking"} strategy \footnote{The use of someone else’s hashtag in order to promote one's own social media agenda}. Our findings suggest that right-wing politicians (and their supporters/retweeters) actively and effectively polarise the discourse not just by using their own party hashtags, but also by \textit{"hashjacking"} the political party hashtags of other established parties. The results underline the necessity to understand the success of right-wing parties, online and in elections, not entirely as a result of external effects (e.g. migration), but as a direct consequence of their digital political communication strategy.\\

\end{abstract}
\section{Introduction}
In recent years many countries were facing increasingly polarised political discourses and social networking and micro-blogging sites play a crucial role in these processes. While the Internet has for long been seen to promote public sphere like spaces for democratic discourse \cite{dahlgren_internet_2005}, the most recent developments remind more of echo chambers, in which particularly the far-right is successful in promoting their worldview \cite{colleoni_echo_2014}, \cite{kramer_populist_2017}, \cite{grinberg_fake_2019}\footnote{With destructive effects on societal trust \cite{braesemann2020between,stephany2019deepens,stephany2020estonia}}. This study investigates the use of political party hashtags on Twitter with a network approach that allows analysing the underlying structures of communication on Twitter with the German far-right party AfD as an example. By using community detection algorithms, we identify different clusters within hashtag discourses and are able to show with a logistic regression model that right-wing users or supporters of the AfD are much more likely to use other party hashtags as part of a strategy called "hashjacking" \cite{darius2019hashjacking,shah_candidate_2015}. By using this "hashjacking" strategy official AfD accounts, that are among the most retweeted users in the sample, aim to increase the reach of individual messages and to co-opt the hashtags of other political parties.

\section{Hypotheses}
This study investigates two main hypotheses based on the background literature:
First, we assume that the retweet networks of German political party hashtags are polarised and moreover, we expect that the retweet network using \#AfD is the most polarised of all retweet networks, for the case of German political party hashtags.

Secondly, we expect an overlap between different hashtag discourses and thus users to appear in multiple retweet networks, because they have used multiple political party hashtags during the observation period. Additionally, we expect the cluster overlaps to be asymmetrical: a certain group of users (AfD support) is more likely to use other party hashtags in opposition (indicator of "hashjacking").

\section{Research Design}
The study is based on Twitter data that was collected by accessing Twitter's Rest API and using political party hashtags of German parties represented in the federal parliament (\#AfD, \#CDU, \#CSU, \#FDP, \#GRUENE, \#LINKE, \#SPD) as a macro-level selection criterion. In total this study builds on a sample (n=173,612) of all public Tweets using one or multiple of the selected political party hashtags between May 28th 00:00:00 (CEST) and June 4th 2018 23:59:59 (CEST) on Twitter. The analysis focuses on a network approach and a visualisation of the retweet networks in Gephi using the Force2 layout algorithm \cite{jacomy2014forceatlas2} for each political party hashtag where retweeting creates a link (edge) between two accounts (nodes).

\begin{figure}[h]
\begin{center}
\includegraphics[width=0.9\textwidth, keepaspectratio]{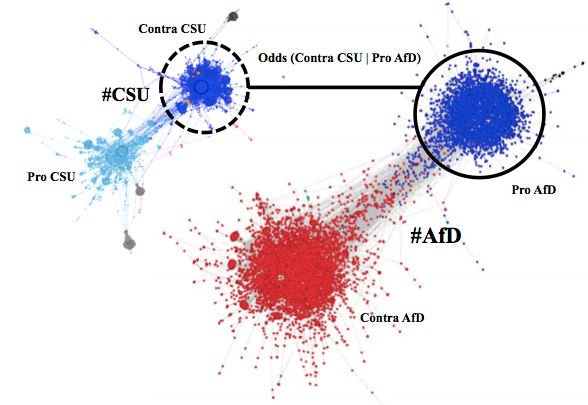}
\end{center}
\caption{\sf{Users of the political party hashtag \#AfD are clearly clustered in two groups (right hand side). Users in the contra-AfD cluster have shared tweets in which \#AfD had been used with negative connotations, while users in the pro-AfD cluster have predominantly shared Tweets in which \#AfD had a positive connotation. All prominent political AfD accounts are located in this cluster. For the hashtags of other parties, like \#CSU, similar clusters emerge (left hand side). We estimate the probability that a user retweeting the CSU-hashtag in critical tweets is also part of cluster A, the pro-AfD cluster: $Odds(X \epsilon Contra CSU | X \epsilon Pro AfD)$.}}
\label{fig:scheme}
\end{figure}

The literature indicated that political discourses on Twitter show polarised or clustered structures due to the retweeting behaviour\footnote{Similar to investigations with online platforms like \textit{Wikipedia} \cite{stephany2017exploration} or \textit{StackOverflow} \cite{stephany2020coding}}. Consequently, the analysis will focus on the retweeting networks of the chosen political party hashtags. 
In a first step of analysis the modularity (community detection) algorithm \cite{blondel2008fast} assigns the nodes to different communities based on the structural properties of the network graph and the cluster membership is indicated by the colour of nodes in Figure \ref{fig:scheme}. Thereafter, the interpretability of the clustering, as being in support of or opposition to a party, is controlled with a qualitative content analysis of the 50 most retweeted accounts similar to Neuendorf et al. \cite{neuendorf_content_2017}. This pro-/contra-polarisation of each party retweet network gives an indication of groups and whether individual nodes from these clusters are likely cluster of other party hashtags. Similarly, we perform a qualitative sentiment analysis of 30 randomly selected retweets (see Table \ref{tab:sentiment}).

Regarding the second hypotheses we assume that a high pro-party X \& contra-party Y association indicates "hashjacking" strategies. Consequently, we use a logistic regression model to test all cluster combinations (as the likelihood to be in a contra-cluster of party Y given a node was in the pro-cluster of party X). We decide to apply a logistic model for the assessment of cross-cluster heterophily, since our dependent variable is binary (location in contra-cluster) and the resulting odds are easy to interpret. 
Assuming there is no group that uses other party hashtags more frequently, users from all clusters should have the same odds to appear in the other network clusters. Thus, a high affiliation between two clusters in terms of their users being more likely to appear in both of them is an indication for strategic hashtag use or "hashjacking".

\section{Results}
The network graphs and modularity assignments confirm the assumption that all retweet networks pertaining to the different political party hashtags are polarised or clustered (the graphs for \#AfD and \#CSU are shown as an example in Figure \ref{fig:scheme}). The assumed pro-party X/contra-party Y alignment was underlined by a qualitative content analysis of the 50 most retweeted accounts in each network, that indicated a high content share of AfD politicians in the Pro-AfD and contra-party X clusters, whereas in the Contra-AfD cluster the most retweeted accounts were journalists, media accounts and outspoken opponents of the AfD (mentioning the opposition in their bio or usernames).

\begin{figure}[!htbp]
\begin{center}
\includegraphics[width=0.9\textwidth]{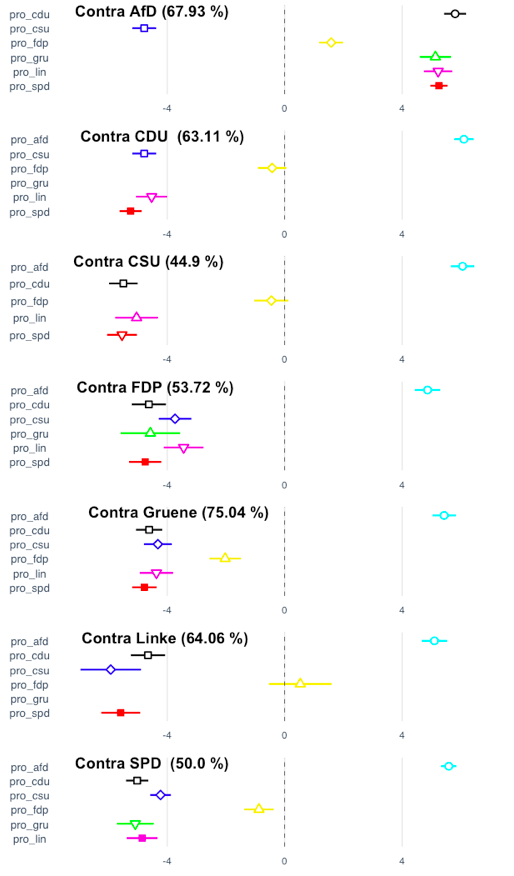}
\end{center}
\caption{\label{fig:coefplot}\sf{Results of the logistic regression models. Dots indicate odds with a 99 percent confidence interval, the relative size of the party contra-cluster is given in parentheses. We find remarkably high association between the pro-AfD cluster and the contra-clusters of all other parties, e.g, users in the contra-CSU cluster are six times more likely to be located in the pro-AfD cluster than users in the contra-CSU cluster on average.}}
\end{figure}

We consult the results of seven logistic regression models to assess the odds for the users in a contra-party cluster to be located in the pro-cluster of the remaining other parties. Figure \ref{fig:coefplot} illustrates the coefficients (odds) of the seven regression models.
The results indicate that users that were located in the pro-AfD cluster are much more likely than users of any other pro-party cluster to appear in the contra-clusters of a given party. When looking at the contra-AfD cluster, however, pro-party users are very likely to be in the contra-cluster of the AfD, which is in line with our expectation of polarised retweeting behaviour corresponding to political party hashtags.

\section{Conclusion}
In this study we understand discourse polarisation as a result and a strategy of far right and right-wing populist parties and are able to show, in the case of the German “Alternative f\"ur Deutschland” (AfD), how official politicians and their supporters polarise the political Twitter discourse by using hashtags strategically in a way that we call "hashjacking". The results underline the assumption of a high polarisation between the AfD and the rest of the political spectrum, whereas it needs to be noted that the FDP seems to be less affected by this polarisation or its supporters seem to restrain from "hashjacking" strategies. While, as expected, we find a high likelihood of AfD supporters to use "hashjacking", we also observe a high likelihood of other parties supporters to use the \#AfD. To assess whether clusters show a high similarity in tweet content a sentiment classifier could be used to verify the structural clustering. Concluding, our study underlines the extent of online polarisation and stresses the importance of awareness for this polarisation when mapping political debates on social media.

\newpage
\section{Appendix}\label{sec:appendix}

\begin{table}[!htbp] \centering 
\caption{\sf{For thirty randomly selected retweets cluster assignment and content sentiment towards AfD are compared. A pearson correlation coefficient of 0.92 indicates that cluster assignment and content's sentiment are well aligned.}}
\label{tab:sentiment}
\scalebox{1}{
\begin{tabular}{@{\extracolsep{5pt}}lcc} 
tweet id            & assigned cluster & sentiment \\
\hline \\[-1.8ex] 
1001006481305690000 & Pro AfD          & +         \\
1001192437107260000 & Anti AfD         & -         \\
1002961768178730000 & Anti AfD         & -         \\
1003348201141720000 & Anti AfD         & -         \\
1002683871828360000 & Pro AfD          & +         \\
1002902201046650000 & Pro AfD          & +         \\
1003603237407190000 & Pro AfD          & +         \\
1001894549344710000 & Anti AfD         & -         \\
1001431016492410000 & Anti AfD         & -         \\
1001068784654910000 & Pro AfD          & +         \\
1001512598989350000 & Anti AfD         & -         \\
1001185575809010000 & Anti AfD         & -         \\
1003495411330480000 & Pro AfD          & ?         \\
1002186342430880000 & Pro AfD          & +         \\
1001940030473030000 & Pro AfD          & ?         \\
1002228287425400000 & Pro AfD          & +         \\
1003273747158220000 & Pro AfD          & +         \\
1002468422477800000 & Pro AfD          & +         \\
1001153121622600000 & Pro AfD          & +         \\
1001034562187780000 & Anti AfD         & -         \\
1002919550197870000 & Anti AfD         & -         \\
1001013938966810000 & Pro AfD          & +         \\
1003239306176220000 & Pro AfD          & ?         \\
1002163520484540000 & Pro AfD          & -         \\
1001754476695420000 & Anti AfD         & ?         \\
1002086300806250000 & Anti AfD         & -         \\
1002370301802440000 & Pro AfD          & ?         \\
1002804299242650000 & Anti AfD         & -         \\
1000957345516610000 & Anti AfD         & -         \\
 \hline \\[-1.8ex] 
\end{tabular}}
\end{table} 

%
%
%
\newpage
\bibliographystyle{splncs04}
\bibliography{Bibliography}
\end{document}